\documentclass[twocolumn,secnumarabic,amssymb, nobibnotes, aps, prd]{revtex4-1}   

\usepackage{amsmath}
\usepackage{amssymb}
\usepackage{graphicx}

\makeatletter
\newenvironment{figurehere}
  {\def\@captype{figure}}
  {}

\makeatother

\begin{document}

\title{A power-law distribution of phase-locking intervals does not imply critical interaction}
\author{M. Botcharova$^{1,2}$}
\author{S.F. Farmer$^2$}
\author{L. Berthouze$^{3,4}$}
\email{L.Berthouze@sussex.ac.uk}
\affiliation{ $^{1}$CoMPLEX - Centre for Mathematics and Physics in the Life Sciences and Experimental Biology}
\affiliation{ $^{2}$Institute of Neurology, UCL, UK}
\affiliation{$^{3}$Centre for Computational Neuroscience and Robotics, University of Sussex, UK}
\affiliation{$^{4}$Institute of Child Health, UCL, UK}


\begin{abstract}
\textbf{Neural synchronisation plays a critical role in information processing, storage and transmission. Characterising the pattern of synchronisation is therefore of great interest. It has recently been suggested that the brain displays broadband criticality based on two measures of synchronisation --  phase locking intervals and global lability of synchronisation -- showing power law statistics at the critical threshold in a classical model of synchronisation. In this paper, we provide evidence that, within the limits of the model selection approach used to ascertain the presence of power law statistics, the pooling of pairwise phase-locking intervals from a non-critically interacting system can produce a distribution that is similarly assessed as being power law. In contrast, the global lability of synchronisation measure is shown to better discriminate critical from non critical interaction.}   
\end{abstract}

\maketitle

\section{Introduction}
The notion of criticality has been hotly discussed in relation to its presence in the human brain \cite{chialvo2004,sornette,timme,stam2012,werner2011}. Support for the concept of a critical brain has emerged from comparing brain dynamics at various scales with the dynamics of physical systems at criticality. Much impetus for this line of work has come from the observation of power laws, a necessary but insufficient condition for criticality, in distributions associated with neuronal avalanches \cite{beggsandplenz, shew}, but further evidence has come from the application of methods from statistical physics for identifying spatio-temporal scaling functions in fMRI \cite{plenzandchialvo,expert}, long-range temporal correlations in amplitude fluctuations of bandpass filtered electro/magneto-encephalogram (M/EEG) \cite{linkenkaer2001,poil2012} as well as universal scaling functions in the activity of individual neurons \cite{friedman, ribeiro}. Functionally, it has been difficult to attribute relevance to these findings other than by making observations of difference in some scaling parameter between different human subject populations or with the subject's age. It would therefore be of great interest to find evidence of criticality in the synchronisation of activity between different brain areas i.e. a parameter that has been directly linked with information processing, storage, and transmission \cite{fries,singer}. 

A system at, or close to, a critical phase transition has been associated with the possibility of rapid reconfigurations in response to external stimuli \cite{shew, werner2010}. Kitzbichler et al. \cite{bullmore,kitzbichler} argue that rapid state changes are crucial for the brain to deal with the environment it meets. They suggest that in some situations, an extensive cognitive effort is required and information transfer needs to be maximised between brain regions, and at other, relatively quiescent periods, the greater concern is minimising neuronal wiring costs \cite{kitzbichler}. A brain at criticality might allow the necessary rapid transitions in functional connectivity to occur quickly \cite{honey}. Werner \cite{werner2010} indicates that a neurophysiological system in a critical state is best able to learn and remember complex logical rules, by adapting its synaptic weights quickly. Meisel et al. \cite{meisel} suggest that local events can spread rapidly through a system in such a state, and that remaining at criticality prevents the spread both from becoming uncontrollably large, or from dying away without effect. A single element hence has the ability to affect the entire system, which may be crucial to processing external stimuli efficiently \cite{chialvo}. 

To assess criticality of synchronisation, Kitzbichler et al. \cite{bullmore} proposed two measures characterising the pattern of synchronisation in a complex system. The first measure is the frequency density of phase locking intervals (PLI), which are defined as the periods of time for which two oscillators differ in their phase by less than a value of $\pi/4$ in modulus. The phase, here, describes where an oscillator is in its cycle, relative to the origin. It evolves in the interval $\left[ -\pi, \pi \right]$ as the oscillator completes an oscillation. The second measure is the frequency density of the change in number of phase locked pairs between successive time points (global lability of synchronisation or GLS). Both measures are derived from a thresholded wavelet-transformed instantaneous phase difference (further introduced in Sections~\ref{sec:scales} and ~\ref{sec:measures}). Kitzbichler et al. validated the PLI and GLS results by showing that in two known models of critical interaction, namely, the Ising model \cite{ising1,ising2} and the Kuramoto Model \cite{kuramoto0,kuramoto1,kuramoto2} (further discussed in Section~\ref{sec:kuramoto}), these measures display power law distributions at the critical threshold but not in a decoupled system \cite{bullmore}. The presence of this power law in the PLI and GLS was determined using a model selection approach \cite{Konishi2007,Claeskens2008} whereby both the power law and alternative models (log-normal and exponential) are fitted and the best model is decided on the basis of the Akaike Information Criterion (formally introduced in Section~\ref{sec:AIC}). 

Whilst it is true that power law statistics of some observable of the system should be evident in a system at criticality \cite{bak,sornette,BakTangWiesenfeld,phillips}, the point has been made that power laws could result from the superposition of multiple processes each with their own characteristic time scale \cite{wagenmakers} or from the use of thresholds \cite{touboul}. Given this, we ask whether power law distributions in the PLI and GLS measures introduced in \cite{bullmore} are uniquely indicative of a system in a critical state. Our approach is to pool the phase locked intervals (respectively, the number of phase locked pairs between successive time points) of a non-critically interacting system of Kuramoto oscillators and compare the resulting distributions with those derived from a critically-coupled system. If this pooling produces distributions that, within the limits of a model selection approach, cannot be distinguished from those of a critically-coupled system then we suggest that this approach to inferring criticality is suspect. 
To do so, we consider a system formed from a collection of independent paired oscillators, which we refer to as the Independent Pairs model. The two oscillators making up a pair are coupled, having phases evolving according to the Kuramoto differential equations (formally introduced in Section~\ref{sec:kuramoto}), but there is no connection between pairs. Each pair can snap into synchronisation at a coupling value unique to itself, however, there is no collective order parameter to unite their progressive synchronisation, i.e., this system can have no critical coupling value. 

The paper is organised as follows. After a brief review of the Kuramoto oscillators (Section~\ref{sec:kuramoto}), we derive analytically the phase difference between two sine-phase coupled oscillators, which makes it possible to generate a large number of Independent Pairs, with natural frequencies drawn from a normal distribution and pair-wise coupling a free parameter (Section~\ref{sec:pairformulation}). After summarising the methodology of Kitzbichler et al.  (Sections~\ref{sec:natfreq}-\ref{sec:AIC}), we compare its application to both the Kuramoto model and our Independent Pairs model (Sections~\ref{sec:Kresults}-\ref{sec:IPMResults}), revealing the coupling parameters under which PLIs and GLSs may give rise to power laws within a model selection approach.

\section{Methods and Materials}

\subsection{The Kuramoto model}\label{sec:kuramoto}
The Kuramoto model is a classical model of synchronisation \cite{acebron,chopra05}. It has been widely used to study the oscillatory behaviour of biological systems such as the sleep and body temperature cycles in humans \cite{pikovsky, strogatz00}, heart pacemaker cell firing \cite{acebron, pikovsky, strogatz00}, neuronal firing \cite{breakspear,bullmore,pikovsky} and fire-fly flashing \cite{acebron,pikovsky, strogatz, strogatz00}.

The Kuramoto model describes the phase behaviour of a system of mutually coupled oscillators with a set of differential equations. Each of $N$ oscillators in the system rotates at its own natural frequency $\left\lbrace \omega_i, i = 1,...,N \right\rbrace $, drawn from some distribution $g(\omega)$. However, it is attracted out of this cycle through coupling $K$, which is globally applied to the system. 
The differential equation to describe the time evolution of the phase $\theta_i$ of oscillator $i$ in such a system is given by \cite{kuramoto0,kuramoto1,kuramoto2}:
\begin{equation}\label{thetadot}
\dot{\theta}_i = \omega_i + \frac{K}{N} \Sigma^{N}_{j = 1} \mbox{sin}(\theta_j  - \theta_i) 
\end{equation}
Kuramoto \cite{kuramoto0} showed that the evolution of any phase $\theta_i$ can be re-expressed using two mean field parameters, which result from the combined effect of all oscillators in the system. Namely, we may say:
\begin{equation}\label{meanfield}
\dot{\theta}_i = \omega_i + K r \mbox{sin}(\psi - \theta_i) 
\end{equation}
where $\psi$ is the mean phase of the oscillators, and $r$ is their phase coherence, so that:
\begin{equation}\label{order}
r e^{i \psi} = \frac{1}{N} \sum^N_{j=1} e^{i \theta_j}
\end{equation}
This crucially indicates that each oscillator is coupled to the others through its relationship with mean field parameters $r$ and $\psi$, so that no single oscillator, or oscillator pair drives the process on their own. The oscillators synchronise at a phase equal to the mean field $\psi$, and $r$ describes the strength of synchronisation, sometimes referred to as the extent of order in the system  \cite{mirollo,bonilla}. When $r=0$, no oscillators are synchronised with each other. When $r=1$, all oscillators are entrained with each other.

It is easy to see that one solution to Equation \ref{meanfield} is $r \equiv 0$ for all time and coupling, leaving each oscillator to evolve independently at its own natural frequency. Using a limit of $N \rightarrow \infty$, some further deductions can be made, including the fact that when the natural frequency distribution $g(\omega)$ is unimodal and symmetric, another solution can be found for $\theta_i$, with $r$ not equivalent to $0$ \cite{kuramoto0}. A critical bifurcation occurs for sufficiently high coupling, resembling a second-order phase transition \cite{miritello} in which the order parameter (here, $r$) leaves zero and grows continuously with coupling \cite{dorfler, mirollo}. The coupling at the bifurcation is referred to as the critical coupling $K_c$ \cite{dorfler}. While the above definition holds for a system of infinite size, for a finite system such as that considered in this paper, the critical coupling can only be approximated by this theoretical value. In Section~\ref{kvalue}, we will provide an operational definition of critical coupling in a finite size system.

\subsection{Analytic Phase Difference for the Independent Pairs Model}
\label{sec:pairformulation}
An independent pair is defined as two coupled oscillators $i$ and $j$ whose phases evolve according to Equation~(\ref{thetadot}), namely: 
\begin{align}\label{phasediff2}
\dot{\theta}_i - \dot{\theta}_j  & = (\omega_i - \omega_j) + \frac{K}{2} \left(   \mbox{sin}(\theta_j - \theta_i)  -   \mbox{sin}(\theta_i - \theta_j) \right)  \nonumber \\
& = (\omega_i - \omega_j) - K \left(    \mbox{sin}(\theta_i - \theta_j) \right)  
\end{align}
Letting $\bigtriangleup_{ij} = \theta_i - \theta_j $ yields:
\begin{align}\label{diffequation}
\dot{\bigtriangleup}_{ij}  & = (\omega_i - \omega_j) - K  \mbox{sin}(\bigtriangleup_{ij})  
\end{align}
This equation has two solutions depending on whether $ K< \mid\omega_i -\omega_j\mid $ or $ K>\mid \omega_i - \omega_j\mid $. If we let $ C = \frac{K}{(\omega_i - \omega_j)}$, and $D$ is an integrating constant, then the solution for $ K< \mid\omega_i - \omega_j\mid $ is:
\footnotesize 
\begin{align}\label{delta}
\begin{split}
\bigtriangleup_{ij} &= 2 \mbox{tan}^{-1} \left[    \left( \sqrt{1 - C^2}\right)\mbox{tan} \left(  \frac{\left( t-D\right) (\omega_i - \omega_j) \sqrt{(1-C^2) } }{2}   \right) \right. \\
& \left. +C\vphantom{\frac{\left( t-D\right) (\omega_i - \omega_j) \sqrt{(1-C^2) } }{2}} \right] 
\end{split}
\end{align}
\normalsize
The solution for $ K> \mid\omega_i - \omega_j\mid $ is:
\footnotesize 
\begin{align}\label{sync}
\bigtriangleup_{ij} &= 2 \mbox{tan}^{-1} \left[  \sqrt{C^2 - 1} \left( \frac{e^{-t (\omega_i - \omega_j) \sqrt{(C^2-1)}} - A}{ A+e^{-t (\omega_i - \omega_j) \sqrt{(C^2-1)}}} \right) + C  \right] 
\end{align}
\normalsize
with $A$ an integrating constant. A full derivation is provided in the Appendix. After deriving this, the authors were made aware that the dynamics of a single pair from this model has previously been described in \cite{adler} in relation to the interaction between a pendulum suspended in a viscous fluid inside a rotating container, and used in \cite{hemmen} as a basis for constructing a Lyapunov function.

The time evolution of $\bigtriangleup_{ij}$ is dependent on two parameters: the coupling $K$, and the difference between the natural frequencies of rotation, $\omega_i - \omega_j$ of the two oscillators. The selection of these two quantities is crucial to further analysis and we look at each in turn. 

\subsection{Natural Frequencies}
\label{sec:natfreq}

The natural frequencies of oscillators in the Kuramoto system considered in \cite{bullmore} were drawn from a normal distribution $ \mathcal{N} \left( 0, 1\right)$. As any normal distribution may be scaled and shifted so that it is equivalent to one with a mean of $0$ and a standard deviation of $1$, we consider that our natural frequencies are also distributed with $\omega_i \sim \mathcal{N} \left( 0, 1\right)$ without loss of generality. If both natural frequencies $\omega_i$ and $\omega_j$ are drawn in this way, then by laws of normal distributions, $\omega_i - \omega_j \sim \mathcal{N} \left( 0, 2 \right) $. As the quantity  $\omega_i - \omega_j$ only is of interest to us in order to calculate $\bigtriangleup_{ij}$ (Equations \ref{delta} and \ref{sync}), we draw values from a distribution of $\mathcal{N} \left( 0, 2 \right) $ for the Independent Pairs Model.  

\subsection{Coupling Parameter}\label{kvalue}

The critical coupling parameter was calculated analytically by Kuramoto under a certain set of assumptions \cite{kuramoto0}. Namely, if the probability distribution of the natural frequencies $g(\omega)$ is unimodal and symmetric, and the number of oscillators is infinite ($N \rightarrow \infty$), then the analytic critical coupling parameter $K_c$ is:

\begin{equation}
K_c = \frac{2}{\pi g(0)}
\end{equation}

And, in the case of $g(\omega) = \mathcal{N} \left( 0, 1\right) $:
\begin{equation}
K_c = \frac{2 \sqrt{2}}{\sqrt{\pi}} \simeq 1.596
\end{equation}

In any feasible realisation of the Kuramoto model, the assumption $N \rightarrow \infty$ is not realistic. This means that the theoretical value of $K_c \simeq 1.596$ is not necessarily the precise coupling parameter for which the system reaches critical behaviour. Kitzbichler and colleagues \cite{bullmore} describe two practical measures characterising the onset of synchronisation with increasing coupling. The first is the change in the `effective mean-field coupling strength', $\Delta (Kr)$. If the value of $Kr$ exceeds the difference between the natural frequency and the mean phase $\omega_i - \psi$ (in modulus) i.e. $\vert \omega_i - \psi \vert <  Kr$, then oscillator $i$ will synchronise to the mean field \cite{mertens}. Thus the value of $K$ at which $Kr$ increases maximally is the coupling value at which the greatest number of oscillators are drawn into the mean field, i.e., a defining feature of the critical point in the system. The second measure is the change in the time-averaged number of synchronised pairs $N_{SP}$ as the coupling increases, $\Delta N_{SP}$. Again, this describes the point at which the greatest change in synchronisation occurs, i.e., the critical point. The two measures $\Delta (Kr)$ and $\Delta N_{SP}$ peak at the same point. We shall call the coupling value at this point the effective critical coupling value for our system. 

In contrast, in our Independent Pairs model, there is no longer a global critical coupling parameter $K_c$ since there can be no mean field. From the two distinct analytical solutions for $\bigtriangleup_{ij}$ (Equations \ref{delta} and \ref{sync}) we see that each pair of oscillators will synchronise independently when $K$ exceeds $\mid\omega_i - \omega_j\mid $ for that pair. Some insight can nevertheless be gained by calculating the measures derived from a standard Kuramoto model, namely, $r$, $N_{SP}$, $\Delta (Kr)$ and $\Delta N_{SP}$. 

\begin{figure*}
\includegraphics[scale=0.4]{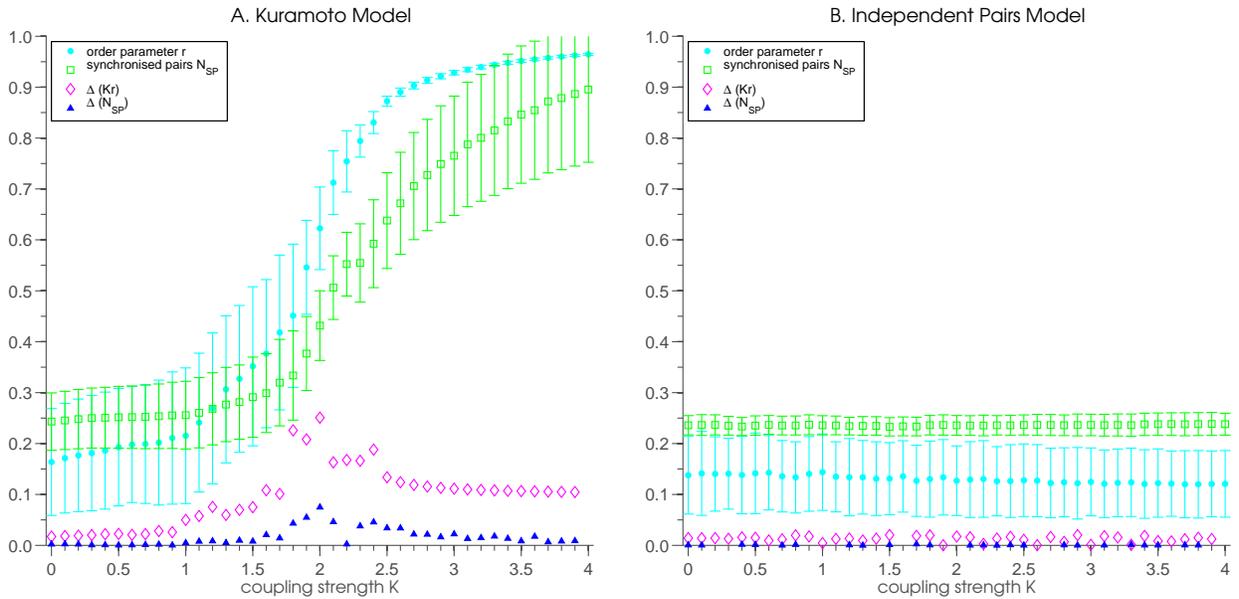}
\caption{Plot A. shows the evolution of order parameter $r$ for the Kuramoto model with cyan solid circles (error bars show standard deviations). The coupling parameter $K$ increases along the $x$-axis. The hollow purple diamonds show $\Delta Kr$, the change in order parameter multiplied by coupling (error bars not shown for readability). The time-averaged number of synchronised pairs, $N_{SP}$ is shown with hollow green squares (error bars show standard deviations), and the difference in $N_{SP}$, $\Delta N_{SP}$, is indicated by solid blue triangles (error bars not shown for readability). The peaks in $\Delta Kr$ and $\Delta N_{SP}$ can be used to indicate the location of the critical point for a specific system, which for this selection of natural frequencies occurs at around $K=2$. This effective coupling value of $K=2$ will be used throughout the paper. Note that, for the Kuramoto model, the order increases with rising coupling. Plot B. displays the corresponding measures $r$, $\Delta Kr$, $N_{SP}$ and $\Delta N_{SP}$ for the Independent Pairs model. There is no change in order parameter with coupling, indicating that the oscillators are not critically coupled to a mean field. \newline\newline} \label{order_parameters}
\end{figure*} 

As shown by Figure \ref{order_parameters}A, there is a clear growth in order in the Kuramoto model, with the parameter beginning near $0$ for low coupling, and increasing to nearly $1$ after the coupling value exceeds $K=3$. The maximum rise in $Kr$ occurs at around $K=2$, which is therefore the effective critical coupling for this system. A similar pattern is traced by $N_{SP}$, with $\Delta N_{SP}$ peaking at around $K=2$. In this paper, we will provide results for the theoretical critical value $K_c \simeq 1.596$ (occasionally referred to as $K_c \simeq 1.6$), as well as for the (above defined) effective critical coupling for our finite system, $K =2$. This latter value is where we might expect power law statistics to be present in the Kuramoto model. The authors have empirically confirmed that as $N$ increases, the effective critical coupling $K$ converges to the theoretical critical coupling $K_c$ (results not shown, but the effective critical coupling is $K=1.8$ for $N=1000$ for example). It should be noted that although the number of oscillators considered here is limited, $44$ oscillators as in \cite{bullmore}, this system still gives rise to $946$ pairwise interactions, which is more substantial. From a neuroscience viewpoint, it could be argued that $44$ oscillators are sufficient for drawing useful conclusions about a neuronal system. For example, the use of a Kuramoto model of $66$ phase oscillators by the authors of \cite{cabral} led to the emergence of slow activity fluctuations consistent with empirically measured functional neural connectivity. Nevertheless, in order to verify our conclusions, we replicated our analysis with $N=1000$ oscillators yielding similar results (not shown but available upon request from the corresponding author). 

With independent pairs, on the other hand, both the order parameter and the number of synchronised pairs remain unchanged across all coupling values, at the values observed for $K=0$ in the Kuramoto model (see Figure \ref{order_parameters}B). This is because, although the pairs individually synchronise with each other, the frequencies at which they synchronise are distributed across the whole range of possible frequencies.

\subsection{Frequency scales}
\label{sec:scales}

An important feature of the findings in \cite{bullmore} is that the critical behaviour of neural activity extends across a number of frequency scales, so that criticality is referred to as being broadband. The decomposition of the phase difference data into several frequency scales is done using a Hilbert wavelet transform, and was implemented computationally here using the algorithms from~\cite{whitcher04,whitcher05,selesnick}. Specifically, wavelet scales $3$ - $11$ were used, corresponding to frequencies of $125-62.5$Hz, $62.5-31$Hz, $31-15.5$Hz, $15.5-8$Hz, $8-4$Hz, $4-2$Hz, $2-1$Hz, $1-0.5$Hz, and $0.5-0.25$Hz. 

First, Kitzbichler et al. \cite{bullmore} construct two signals denoted $s_i$ and $s_j$ hereafter, by taking the cosine of phases  $\theta_i$ and $\theta_j$ respectively. They then take the $k$-th scale wavelet transforms of $s_i$ and $s_j$ to obtain $\mathcal{W}_k (s_i)$ and $\mathcal{W}_k (s_j)$, which are time-varying complex vectors of wavelet coefficients. Each set of wavelet coefficients quantifies the power of the signal in the corresponding frequency band. These two sets of wavelet coefficients are multiplied element-wise to form the vector $\mathcal{W}_k (s_i)^{\dagger} \mathcal{W}_k (s_j)$, where the symbol $\dagger$ indicates the complex conjugate. This vector is then normalised by dividing it (again, element-wise) by the element-wise product $\mid \mathcal{W}_k (s_i)\mid \mid \mathcal{W}_k (s_j) \mid$ where operator $\mid . \mid $ denotes the modulus. The result is an instantaneous time-varying complex phase vector:

\begin{equation}\label{Cij}
C^k_{ij} =  \frac{ \mathcal{W}_k (s_i)^{\dagger} \mathcal{W}_k (s_j) }{ \mid \mathcal{W}_k (s_i)\mid \mid \mathcal{W}_k (s_j) \mid }  
\end{equation}
 
To ensure a more robust and less noisy estimate of the phase relation, the instantaneous phase vector is smoothed by using a moving average of the numerator and the two vectors contributing to the denominator of $C^k_{ij}$, yielding a new vector $\bar{C}^k_{ij}$ given by:
\begin{equation}
\bar{C}^k_{ij} =  \frac{ \langle \mathcal{W}_k (s_i)^{\dagger} \mathcal{W}_k (s_j) \rangle }{ \sqrt{   \langle\mid \mathcal{W}_k (s_i)\mid ^2 \rangle   \langle \mid \mathcal{W}_k (s_j) \mid ^2 \rangle }  }
\end{equation}
Here the operator $\left\langle . \right\rangle $ denotes that a moving average is taken. The length of the sliding window used for the moving average is set to the number of time steps spanning 8 oscillation cycles at the highest frequency in that wavelet scale \cite{bullmore}. 

The argument of $\bar{C}^k_{ij}$ is then taken as a measure of the phase relationship of the two oscillators $i$ and $j$ corresponding to wavelet scale $k$, so that $ \bigtriangleup^k_{ij} = \mbox{arg}(\bar{C}^k_{ij} )$. 

In the Independent Pairs model, the phase differences within each pair are known analytically (see Section~\ref{sec:pairformulation}), however, they are not associated with particular wavelet scales. To produce probability distributions comparable to those in~\cite{bullmore}, surrogate pairs of signals were created with the first signal evolving constantly at a frequency given by a base value drawn from the distribution of natural frequencies $g(\omega)$, and the second signal phase shifted from the first by $\bigtriangleup^k_{ij}$.

\subsection{PLI and GLS}
\label{sec:measures}
In this section, we will use $\bigtriangleup^k_{ij}(t)$ to denote the value of $\bigtriangleup^k_{ij}$ at time $t$. For phase difference $\bigtriangleup^k_{ij}$ between two oscillators $i$ and $j$, the PLIs are defined as the duration (in seconds) for which $-\alpha < \bigtriangleup^k_{ij} (t) < \alpha$, for some threshold $\alpha$. This definition was given by \cite{bullmore} with $\alpha = \pi/4$. 

The GLS was also defined in \cite{bullmore} and characterises the evolution of the number of synchronised pairs, $\mbox{N}_{SP}$, to describe the lability of synchronisation. The number of synchronised pairs at wavelet scale $k$ is formally defined as:
\begin{equation} 
N_{SP}^k(t) = \sum_{i<j} \left\lbrace \mid \bigtriangleup^k_{ij}(t)\mid < \alpha \mbox{  and  } {M^k_{ij}}^2 (t) > \frac{1}{2} \right\rbrace 
\end{equation}
where ${M^k_{ij}}^2  = \mid \bar{C}^k_{ij}  \mid ^2$ is proposed as a measure of the significance of the phase difference estimate $\bar{C}^k_{ij}$, and $\alpha = \pi/4$ as above. It should be noted here that the condition ${M^k_{ij}}^2(t) > \frac{1}{2}$ introduces an additional threshold. The use of thresholds on otherwise stochastic data has been shown by Touboul et al. \cite{touboul} to occasionally give rise to spurious power laws. 

The GLS at scale $k$ is then obtained by calculating the square of the difference in the number of phase-locked pairs between two successive points in time:
\begin{equation}
GLS^k = \mid  N_{SP}^k(t+\delta t) - N_{SP}^k(t) \mid ^2
\end{equation}
where $\delta t$ is an increment in time and $k$ denotes the wavelet scale. 

From examination of our analytic equations for phase difference (Equations \ref{delta} and \ref{sync}), we observe that the phase difference $\bigtriangleup^k_{ij}$ changes with time in a very structured way. For $ K<\mid\omega_i -\omega_j\mid$, $\bigtriangleup^k_{ij}$ is a periodic function. For $K>\mid \omega_i - \omega_j\mid $, there is a short-lived transient before $\bigtriangleup^k_{ij}$ settles to a constant. 

Before we proceed to pool our probability distributions across many pairs of oscillators, we first consider what we might expect from a single pair. 

For $ K<\mid\omega_i -\omega_j\mid$, the lengths of PLIs between two oscillators would be identical within any given oscillation cycle, and the probability distribution will only contain one value. If a given simulation is cut off before a full cycle is complete, or more precisely, before a phase locked interval has come to an end, this may give rise to a second phase locked interval, and the probability distribution may have more than one value in this case. For $K>\mid \omega_i - \omega_j\mid $, the phase difference will be a single constant, either occurring during the transient, or at the permanent value to which the phase difference converges, depending on the starting phase difference, and the value of the final constant. Again, the probability distribution contains one value. 

The GLS can either take the value $1$ if the oscillators either go from being non-phase-locked to phase locked, or the value $0$ if no change occurs. This allows two possible values in the probability distribution. 

For a single oscillator pair, we would therefore not expect to find a valid probability distribution of either PLIs of GLS for any coupling $K$.

This is a trivial, but important point to make. If a single pair of oscillators could give rise to a probability distribution which appeared linear on a log-log plot (as a power law does) for some pairwise coupling value that could be considered `critical' over some small range of values, then the final, observed power law created by pooling many pairs may be the result of a simple superimposition of these smaller linear components. We now demonstrate that the power law could result from a process that does not involve `critical' interactions for any reasonable definition of the term (even on a pairwise level), but through completely independent systems evolving with no connections between the elements that combine to produce the power law.

\subsection{Akaike Information Criterion}
\label{sec:AIC}
As in~\cite{bullmore}, the presence of power law statistics is assessed using a model selection approach whereby the Akaike's Information Criterion \cite{akaike} is used to compare the goodness-of-fit of a power law distribution with that of two alternative distributions, namely, the exponential and log-normal distributions. It is important to stress that the Akaike Information Criterion only provides a means of comparing models, but gives no information on how good the model is objectively at fitting the data. This means that only the relative values of this measure, for different models, are important. 

For a model using $k$ parameters, with likelihood function $L$, the Akaike Information Criterion is calculated using the following expression:

$$ AIC = 2k - 2 \mbox{ln}(L) $$

As in \cite{bullmore}, this measure was adjusted to account for small sample sizes, using the following:

$$ AIC_{c} = AIC + \frac{2k(k+1)}{n-k-1}   $$ 

where $n$ is the number of observations of the data. This is especially relevant because all three models were fitted to the binned histogram heights, rather than the full data set. Since the basis of the AIC is a log-likelihood function, it can be used with binned data in this way \cite{cowan}. The number of bins used will affect the raw values of the $AIC$, but not the relative values obtained for the models used, so that the best-fitting model will pertain for the data analysed.


\section{Results}
\subsection{Independent Pair model simulation}
\label{sec:IPM}

We simulated pairs of Kuramoto-coupled oscillators alongside our analytic solution. Both were calculated over $1000$ seconds, with an integration time step of $ \delta t = 2^{-11} $ for the simulated oscillators. This provided a total of $1000 \times 2^{11}$ time steps. We then down-sampled the resulting time series by a factor of 2 to obtain a time series with sampling frequency of $2^{10} \mbox{Hz}$. The analytic signal was also generated with a sampling frequency of $2^{10} \mbox{Hz}$. The coupling $K$ was incremented between $0$ and $4$, in intervals of $0.2$, and the two curves were compared.

\begin{figure}
\centering
\includegraphics[scale=0.37]{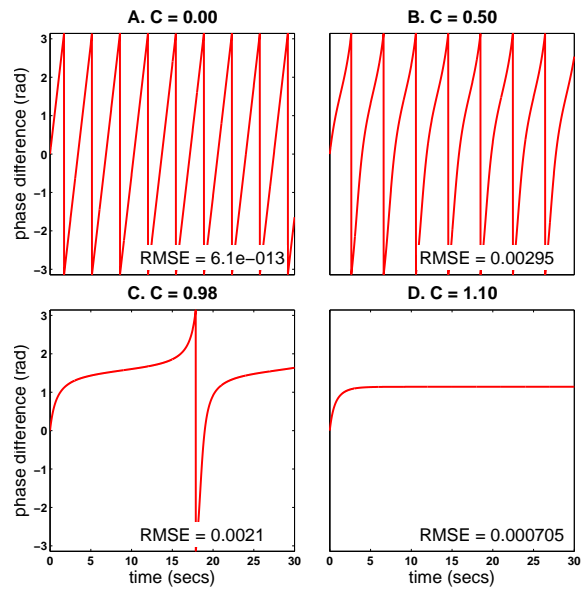}
\caption{ The evolution of phase difference between the oscillators in a two-oscillator Kuramoto system, plotted using our analytic expression (blue), and a simulation of the Kuramoto model by Euler's method (red). The two phase calculations are perfectly superimposed. The root mean square error (RMSE) is shown for different coupling values,for a single simulation. Panels A,B,C have $C<1$ (where $C$ is defined in Section~\ref{sec:pairformulation}), but coupling is increased progressively. The phase evolves periodically. Panel D is the same pair of oscillators, but for $C>1$. There is a brief transient before the oscillators fully synchronise with a constant level of phase difference. The initial phase separation has been set to $\bigtriangleup = 0$ without loss of generality. \newline\newline} \label{phases}
\end{figure}

The behaviour of the phase difference is qualitatively different in the cases $C = \frac{K}{(\omega_i - \omega_k)} <1$ and $C>1$. We demonstrate the phase difference between two oscillators in Figure \ref{phases} as obtained with our analytic expressions alongside a simulation of the Kuramoto model, using Euler's method to iteratively update the phase by Equation \ref{thetadot}.  The two phase calculations are perfectly superimposed. 

Although the root mean square error (RMSE) varies for different coupling values, the normalised RMSE is less than $0.1\%$ for the range of coupling values considered in this paper, demonstrating good agreement between simulated and analytic results.    

It is evident that when the coupling supersedes the difference in natural frequencies ($C>1$), the two oscillators synchronise in exponential time. When the coupling is small ($C<1$), however, the phase difference grows (or falls) at a rate dictated by the frequency difference, but with increasingly lengthy periods of constant phase difference, or synchronisation.  

\subsection{PLI and GLS of Kuramoto model}
\label{sec:Kresults}
\begin{figure*}
\centering
\includegraphics[scale=0.28]{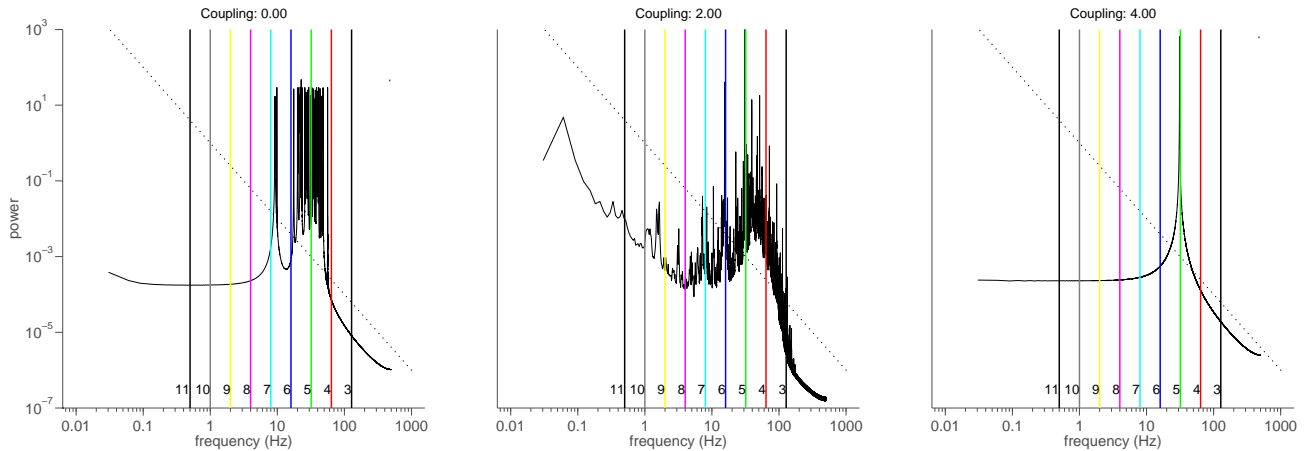}
\caption{Power spectra for a system of 44 Kuramoto oscillators, with natural frequencies drawn from a $\mathcal{N} \left( 60 \pi, 20 \pi \right)$ distribution and three distinct levels of coupling - A) $K = 0$, B) $K = 2$, the effective critical coupling for this specific finite Kuramoto system, as seen from Figure \ref{order_parameters}A and C) $K = 4$. The vertical numbered lines represent wavelet scales $3-11$.} \label{powerspec}
\end{figure*}

As a baseline for comparison, the results of  Kitzbichler et al. \cite{bullmore} on the Kuramoto model were replicated using our own code in the Matlab environment. A system of 44 Kuramoto oscillators, each with a natural frequency drawn from a normal distribution $\mathcal{N} \left( 60 \pi, 20 \pi \right)$, was simulated using the same simulation parameters as in Section \ref{sec:IPM}. We present three different regimes (uncoupled, critically coupled, and super-critically coupled), which yield the power spectra shown in Figure~\ref{powerspec}. 

\begin{figure}
\centering
\includegraphics[scale=0.28]{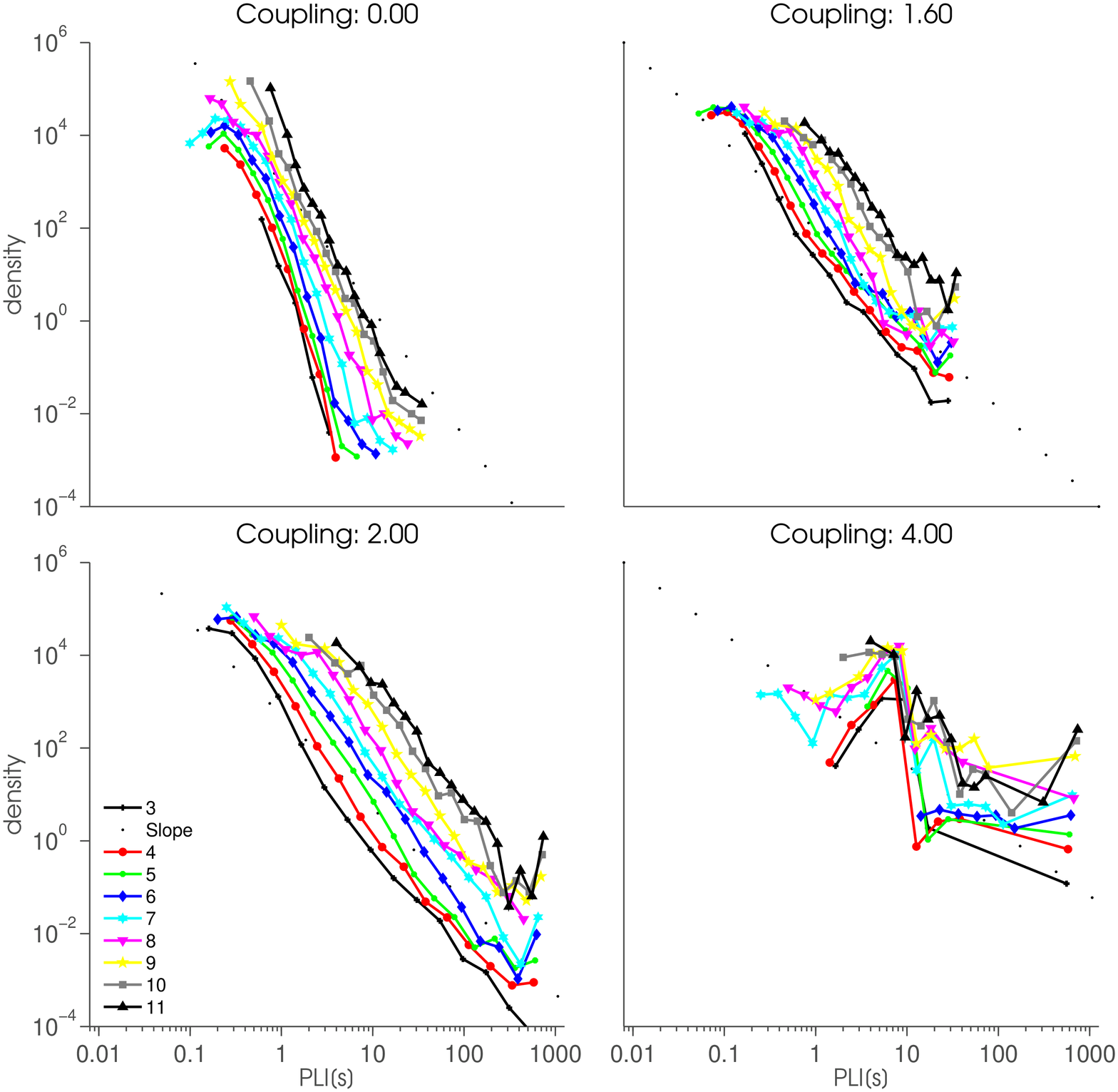}
\caption{Distribution of PLIs in a system of 44 Kuramoto oscillators, with natural frequencies drawn from a $\mathcal{N} \left( 0, 1\right)$ distribution and four levels of coupling - $K = 0$, $K = K_c \simeq 1.6$, $K = 2$ and  $K = 4$ (from top-left, clock-wise). A power law of exponent -2 is shown by a dotted black line. The coloured lines represent wavelet scales $3-11$ (see key).\newline} \label{k_pli_results}
\end{figure}

Next, using 44 oscillators whose natural frequencies were drawn from a $\mathcal{N} \left( 0, 1\right)$ distribution, the PLI and GLS probability distributions were calculated for the following coupling values - $K = 0$, $K = K_c = 1.596$, $K = 2$ and $K = 4$. At $t=0$, all oscillators had a phase $\theta_i=0$. The data presented in figures \ref{powerspec}, \ref{k_pli_results} and \ref{k_gls_results} were obtained from a single run of the model, however, it was confirmed that the results were not sensitive to the exact values of the natural frequencies. 

A histogram for the PLI data was constructed using $20$ logarithmically spaced bins, with the first bin beginning at a single time step of $2^{-10}$ seconds, and the largest bin ending at the total length of the data, of $1000$ seconds. The histogram was then scaled so that each bin count was divided by the total number of PLIs, and then by the bin size that it represented. 

For GLS, we took $1000$ logarithmically spaced bins ranging from a value of $1$ to $10^{4.5}$, as displayed on the plot. The GLS histogram was also scaled. Here each bin count was divided by the total number of counts (sum of all bin counts), and then by the bin size that it represented.

\begin{table}[ht]
\caption{Akaike Information Criterion values for various models applied to the PLI distributions of the Kuramoto model at $K = 2$, the effective critical coupling value for our system. Smaller values indicate a better fit, but comparisons are only meaningful across rows. The smallest value in each row is indicated with an asterisk.}  
\centering 
\begin{tabular}{c c c c c c c} 
\hline\hline 
Wavelet Scale & Power-Law &     & Exponential&      &Log-Normal&           \\ [0.5ex] 
\hline 
3	&	251.04	&	 	&	288.75	&	 	&	116.26	&	$\ast$	\\
4	&	253.87	&	 	&	289.35	&	 	&	123.10	&	$\ast$	\\
5	&	257.03	&	 	&	316.55	&	 	&	157.24	&	$\ast$	\\
6	&	258.62	&	 	&	370.14	&	 	&	218.44	&	$\ast$	\\
7	&	254.59	&	 	&	396.20	&	 	&	252.47	&	$\ast$	\\
8	&	245.74	&	$\ast$	&	359.41	&	 	&	250.97	&	 	\\
9	&	220.50	&	$\ast$	&	343.30	&	 	&	227.93	&	 	\\
10	&	224.56	&	$\ast$	&	318.80	&	 	&	229.26	&	 	\\
11	&	220.38	&	$\ast$	&	306.27	&	 	&	223.93	&	 	\\
\hline
\hline
\end{tabular}
\label{aic_k_pli} 
\end{table}

\begin{table}[ht]
\caption{Akaike Information Criterion values for various models applied to the GLS distributions of the Kuramoto model at $K = 2$, the effective critical coupling value for our system. Smaller values indicate a better fit, but comparisons are only meaningful across rows.  The smallest value in each row is indicated with an asterisk.}  
\centering 
\begin{tabular}{c c c c c c c} 
\hline\hline 
Wavelet Scale & Power-Law &     & Exponential&      &Log-Normal&           \\ [0.5ex] 
\hline 
3	&	-2533.43	&	$\ast$	&	-1019.49	&	 	&	-2478.83	&	 	\\
4	&	-2531.41	&	$\ast$	&	-1296.02	&	 	&	-2484.28	&	 	\\
5	&	-2540.75	&	$\ast$	&	-1351.52	&	 	&	-2490.46	&	 	\\
6	&	-2520.30	&	$\ast$	&	-1304.60	&	 	&	-2473.17	&	 	\\
7	&	-2439.44	&	 	&	-1293.77	&	 	&	-2465.53	&	$\ast$	\\
8	&	-2415.82	&	 	&	-1163.59	&	 	&	-2426.63	&	$\ast$	\\
9	&	-2000.55	&	$\ast$	&	-941.78	&	 	&	-1985.62	&	 	\\
10	&	-1536.79	&	$\ast$	&	-686.48	&	 	&	-1515.75	&	 	\\
11	&	-546.67	&	 	&	-239.38	&	 	&	-568.82	&	$\ast$	\\
\hline
\hline
\end{tabular}
\label{aic_k_gls} 
\end{table}

The Akaike Information Criterion $(AIC)$ was calculated for both the PLI and GLS distributions for all studied coupling values. Only PLI intervals of length $0.1$ seconds or more were used for model-fitting, and these only are shown in the plot. The power-law model was fitted using the procedure described by Clauset et al. \cite{clauset2009}, and implemented using their freely available code, and a minimum data value of $0.1$ seconds. The log-normal and exponential distributions were both fitted using in-built Matlab functions. 

The values obtained for the effective critical coupling $K=2$ are shown in Table \ref{aic_k_pli} for PLIs and Table \ref{aic_k_gls} for GLS. As in \cite{bullmore}, the power law distribution was only found to be the best fit at certain wavelet scales. The $AIC$ values in Table 1 of Kitzbichler et al. \cite{bullmore}, stated as being at critically coupled Kuramoto, favour a power law model of the PLI frequency distribution for $5$ of $9$ wavelet scales, although no value is reported for wavelet scale $11$. 

\begin{figurehere}
\centering
\includegraphics[scale=0.28]{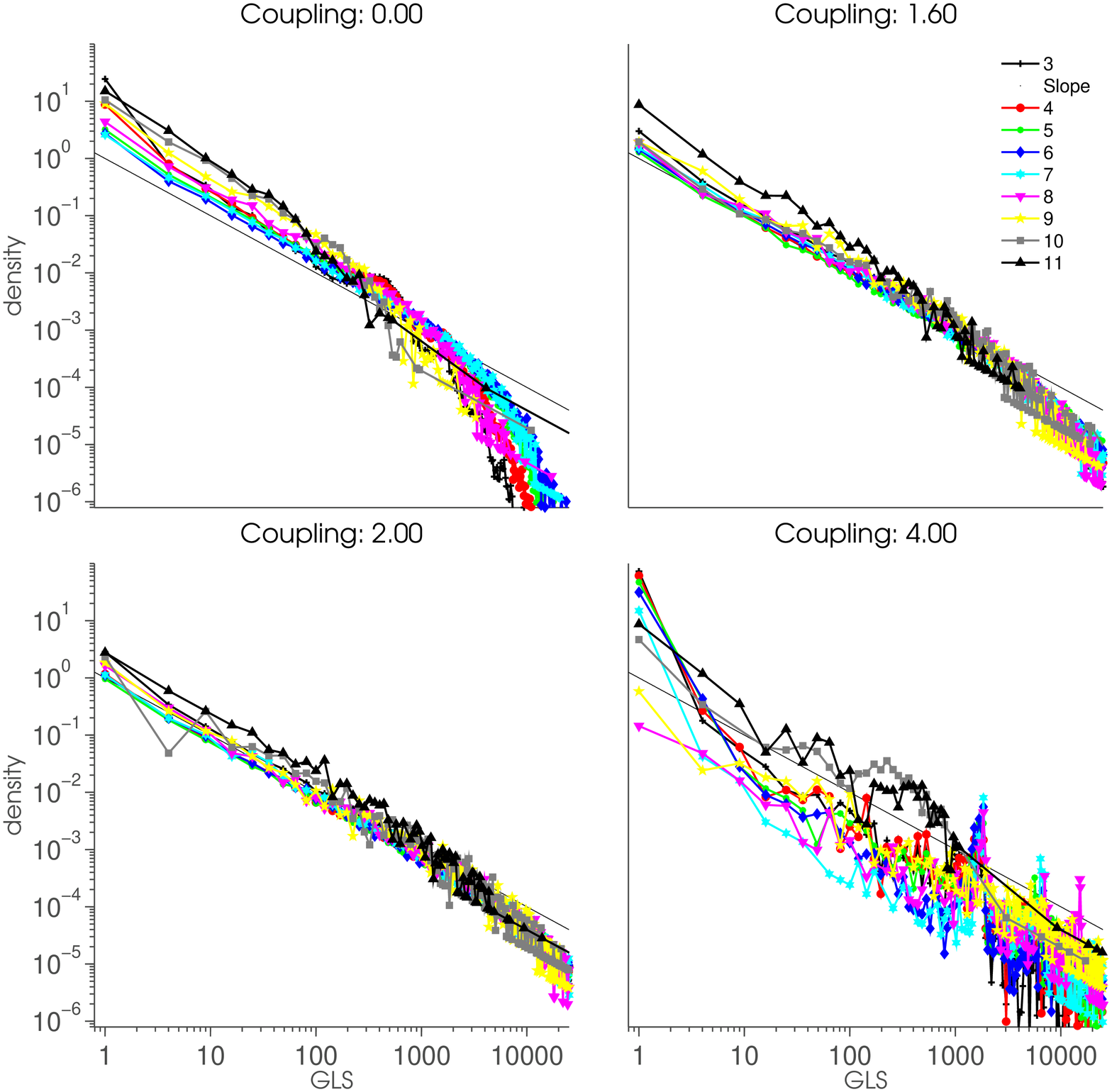}
\caption{ Distribution of GLS in a system of 44 Kuramoto oscillators, with natural frequencies drawn from a $\mathcal{N} \left( 0, 1\right)$ distribution and four levels of coupling - $K = 0$, $K = K_c \simeq 1.6$, $K = 2$ and  $K = 4$ (from top-left, clock-wise). A power law of exponent -1 is shown by a dotted black line. The coloured lines represent wavelet scales $3-11$ (see key).\newline} \label{k_gls_results}
\end{figurehere}

In our system, at the effective critical coupling $K=2$, the power law distribution was the best model for the data for $4$ out of $9$ wavelet scales for the PLI data. Note that the same number of wavelet scales were also best fitted by a power law distribution for coupling values $K=1$, $K=3$ and $K=4$. At coupling $K = K_c = 1.596$, $3$ wavelet scales were best fitted by a power law, and at no coupling, i.e., $K=0$, only $2$ wavelet scales. The log-normal distribution was otherwise the best fit at all coupling values and all other scales. The fact that less than half of the wavelet scales were best fitted by a power law distribution at the critical coupling, combined with the fact that non-critical coupling parameters ($K=1,3,4$) resulted in the same proportion of scales being best fitted by a power law distribution, leads us to conclude that the distribution of PLIs is not a reliable measure of criticality in a finite size Kuramoto system. 

For the GLS probability distribution the coupling values giving greatest resemblance to power law distributions were $K=K_c \simeq 1.6$ and also $K=3$, both with $8$ of $9$ wavelet scales best fitted by the power law model. (The $AIC$ values for the GLS distribution were not included in \cite{bullmore}). In contrast, a power law model was best-fitting for only $2$ wavelet scales at coupling value of $K=0$. It was the best fit for $4$ wavelet scales at coupling $K=1$, for $6$ wavelet scales at coupling $K=2$ and for $3$ wavelet scales at coupling $K=4$. The remaining wavelet scales for all coupling values were again best fitted by a log-normal distribution. The prevalence of good power law fits in the GLS probability distribution across wavelet scales for coupling values $K=K_c$, $2$ and $3$, and the fact that power law distributions were not a good fit for the data resulting from coupling values $K=0$ and $K=4$, collectively suggest that the GLS measure may be an acceptable but not very sensitive indicator of the region of critical coupling for the finite size Kuramoto system.

The probability distributions of PLIs and GLS in Figures~\ref{k_pli_results} and ~\ref{k_gls_results} are consistent with those shown in Figure 3 of \cite{bullmore} for the zero and critical coupling values. For $K=0$, the probability distribution of the PLIs has a drop-off for PLI values above $10^0$. However, our plot at this value differs from that in Kitzbichler et al. \cite{bullmore}, which shows that no intermediate length PLIs exist for many of the scales. We observe PLIs of all lengths from $0.1$ to over $100$ seconds with non-zero probability. We suspect that their data was truncated for display, but no detail is given in the paper. 
The distributions at all wavelet scales appear linear in the log-log space both at theoretical critical coupling of $K_c \simeq 1.6$, and at $K=2$, the effective coupling parameter for this simulation of the Kuramoto system. The range in which this linearity holds is similar to that in \cite{bullmore}, lying between $10^0$ and $10^2$. Our results for coupling values beyond criticality show that the distributions remain power-law-like as the coupling is increased to $K=3$, suggesting that linearity in the log-log space is not specific to $K=K_c$ for this system. This linearity in the log-log space vanishes for $K=4$, where sufficiently many oscillators have synchronised at the mean field phase for the system, which induces a particular interval of phase-locking, indicated by the peak in the distribution. Qualitatively similar observations can be made regarding the GLS distributions. 

\subsection{PLI and GLS in the Independent Pairs model}
\label{sec:IPMResults}

PLI and GLS probability distributions were computed from the phase difference of $1000$ pairs of oscillators with $\omega_i - \omega_k \sim \mathcal{N} \left( 0, 2 \right)$. The length of data, and time steps used were identical to those described in Section \ref{sec:IPM}. The number of pairs was set to a value close to that of the total number ($946$) of pairings available in a system of 44 oscillators. We computed all PLIs across these pairings, and the measures of GLS for all consecutive time points. Histograms of PLI and GLS, and $AIC$ values were computed exactly as in the previous Section (see Figures~\ref{pli_results} and~\ref{gls_results}, and Tables~\ref{aic_m_pli} and~\ref{aic_m_gls}).

\subsubsection{PLI probability distribution}
\begin{figure}
\centering
\includegraphics[scale=0.28]{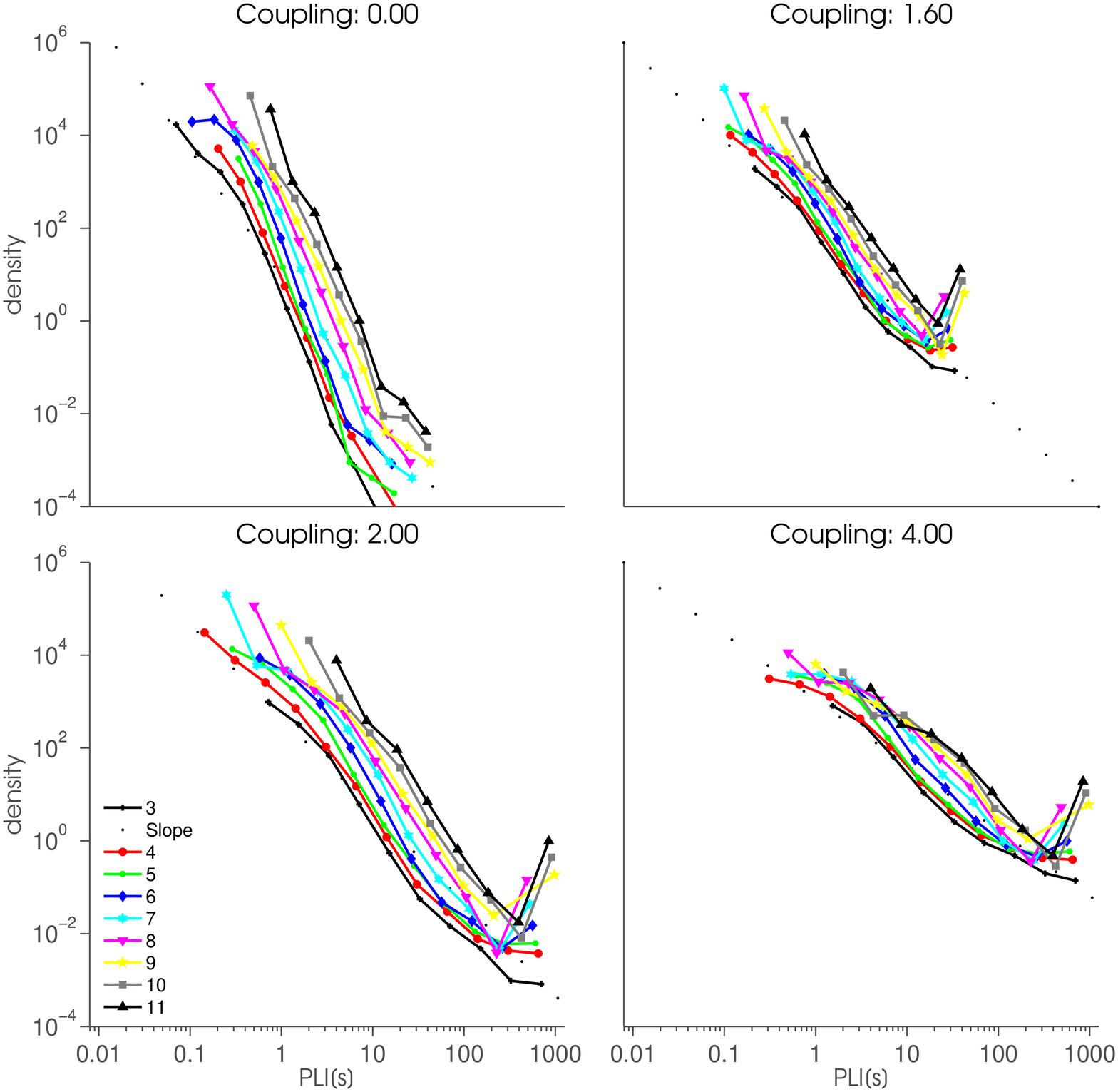}
\caption{Distribution of PLIs in the Independent Pairs Model, with natural frequencies drawn from a $\mathcal{N} \left( 0, 1\right)$ distribution and four levels of coupling - $K = 0$, $K = K_c \simeq 1.6$, $K = 2$ and  $K = 4$ (from top-left, clock-wise). A power law of exponent -2 is shown by a dotted black line. The coloured lines represent wavelet scales $3-11$ (see key).\newline} \label{pli_results}
\end{figure}

As indicated by Figure \ref{pli_results}, the structure of the probability distribution alters as the coupling increases. For $K =0$, there is a drop-off below the power law of the distribution for values of the PLI above $1$ second. At or around the theoretical and effective critical couplings, the log-log plot of the distribution approaches the same power law with slope $-2$ as indicated by \cite{bullmore}. For values up to $K=3$, there is no significant difference between the evolution of PLI probability distributions with coupling in the Independent Pairs model and that of the Kuramoto model. The main dissimilarity arises from the continuing presence of an apparent power law distribution in the `super-critical' range of $K=4$. In the Independent Pairs model, the log-log plot of the distribution retains some of its linearity whereas there is synchronisation to the mean field in the Kuramoto model, as evidenced by a well-defined peak in Figure~\ref{k_pli_results}.

For the Independent Pairs Model, the $AIC$ indicated that the power law distribution best fitted the PLI probability distribution for $4$ of the $9$ wavelet scales, at critical coupling value $K \simeq 1.6$, as well as for coupling values $K=1$ and $K=4$. Both the effective critical coupling value $K=2$ (see Table \ref{aic_m_pli}) and $K=3$ favoured the power distribution for $5$ wavelet scales in contrast to only $1$ wavelet scale for coupling $K=0$. The remaining wavelet scales at all coupling values were best fitted by a log-normal distribution. As there is little difference between the numbers of wavelet scales best fitted by a power law distribution for corresponding coupling values of the Kuramoto and Independent Pairs models, we conclude that the PLI measure is therefore unable to distinguish between critically and non-critically coupled systems. 

\begin{table}[ht]
\caption{Akaike Information Criterion values for various models applied to the PLI distributions of the Independent Pairs Model at $K = 2$, the effective critical coupling value for our system. Smaller values indicate a better fit, but comparisons are only meaningful across rows. The smallest value in each row is indicated with an asterisk.}  
\centering 
\begin{tabular}{c c c c c c c}
\hline\hline 
Wavelet Scale & Power-Law &     & Exponential&      &Log-Normal&           \\ [0.5ex] 
\hline 
3	&	205.74	&	 	&	121.02	&	 	&	49.49	&	$\ast$	\\
4	&	189.05	&	 	&	222.37	&	 	&	120.70	&	$\ast$	\\
5	&	171.14	&	 	&	192.08	&	 	&	107.80	&	$\ast$	\\
6	&	154.09	&	 	&	166.67	&	 	&	93.89	&	$\ast$	\\
7	&	138.37	&	$\ast$	&	241.74	&	 	&	139.03	&	 	\\
8	&	122.33	&	$\ast$	&	210.90	&	 	&	124.66	&	 	\\
9	&	104.09	&	$\ast$	&	174.94	&	 	&	109.51	&	 	\\
10	&	88.21	&	$\ast$	&	161.30	&	 	&	93.26	&	 	\\
11	&	72.94	&	$\ast$	&	129.74	&	 	&	80.59	&	 	\\
\hline
\hline
\end{tabular}
\label{aic_m_pli} 
\end{table}

\subsubsection{GLS probability distribution}

\begin{figure}
\centering
\includegraphics[scale=0.28] {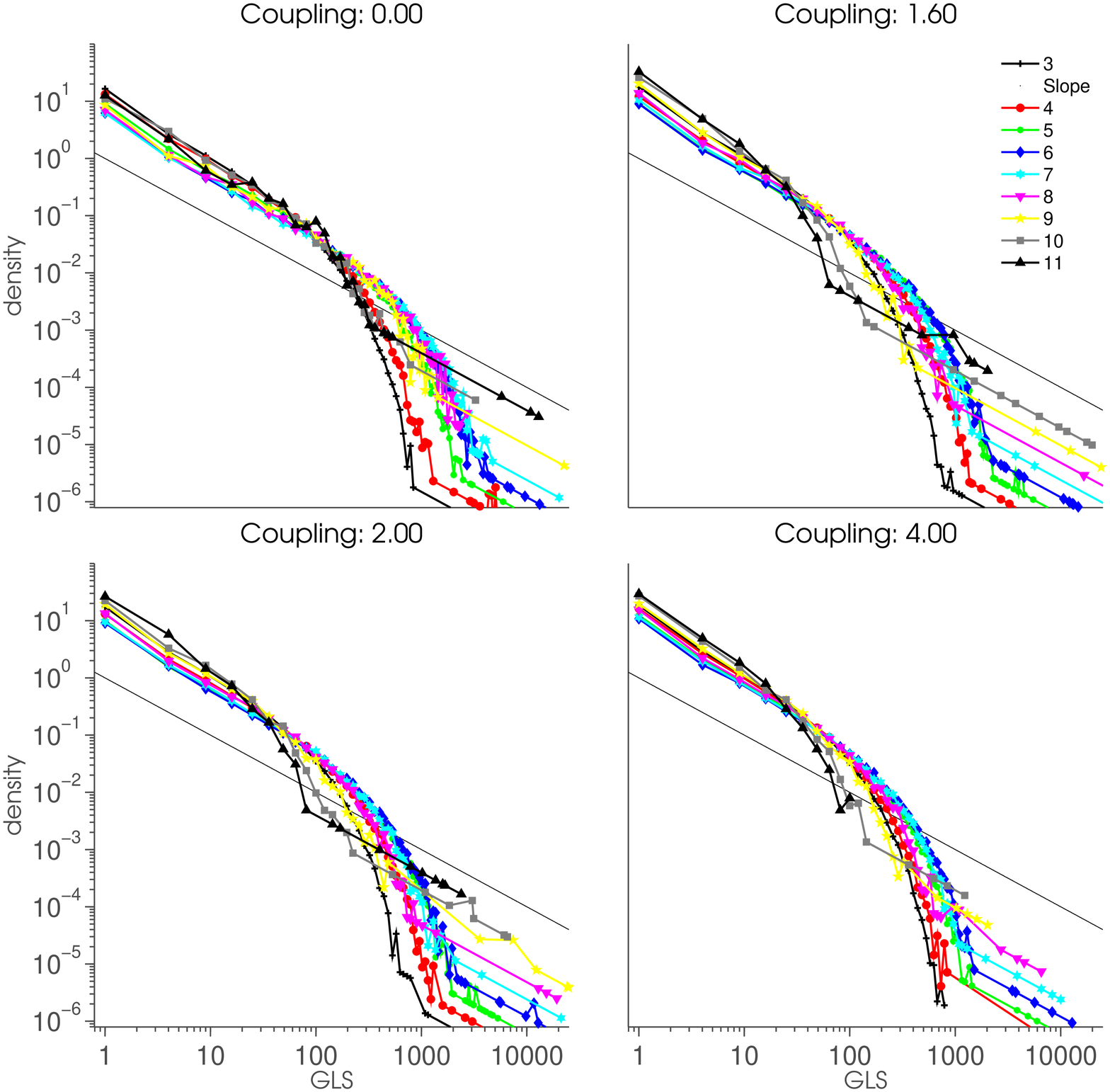}
\caption{Distribution of GLS in the Independent Pairs Model, with natural frequencies drawn from a $\mathcal{N} \left( 0, 1\right)$ distribution and four levels of coupling - $K = 0$, $K = K_c \simeq 1.6$, $K = 2$ and  $K = 4$ (from top-left, clock-wise). A power law of exponent -1 is shown by a dotted black line. The coloured lines represent wavelet scales $3-11$ (see key).\newline} \label{gls_results}
\end{figure}

In contrast to the PLI results, the probability distribution for the GLS of the Independent Pairs model remains largely unaltered as coupling increases, as shown in Figure \ref{gls_results}. The GLS distributions do not resemble those of the Kuramoto model. The range in which the log-log plot of the distribution is linear is narrower with a drop-off in the distribution for values of GLS above $100$s, suggesting that the Global Lability of Synchronisation measure may be more sensitive to the lack of critical interaction in the system. 

For GLS, only $2$ wavelet scales were best modelled by the power law model at the effective critical coupling $K=2$ (see Table \ref{aic_m_gls} for $K=K_c$). $1$ wavelet scale was best fitted by a power law at coupling $K=0$, $3$ at $K=1$, $2$ at $K=K_c$, $4$ at $K=3$, and $3$ at $K=4$. The remaining wavelet scales at all coupling values were best fitted by a log-normal distribution. There is no evident pattern of increasing similarity to a power law of the GLS distribution, as the coupling increases.   

\begin{table}[ht]
\caption{Akaike Information Criterion values for various models applied to the GLS distributions of the Independent Pairs model at $K = 2$, the effective critical coupling value for our system. Smaller values indicate a better fit, but comparisons are only meaningful across rows. The smallest value in each row is indicated with an asterisk.}  
\centering 
\begin{tabular}{c c c c c c c} 
\hline\hline 
Wavelet Scale & Power-Law &     & Exponential&      &Log-Normal&           \\ [0.5ex] 
\hline 
3	&	-297.16	&	 	&	42.78	&	 	&	-301.51	&	$\ast$	\\
4	&	-379.92	&	 	&	8.93	&	 	&	-391.39	&	$\ast$	\\
5	&	-591.87	&	 	&	-54.62	&	 	&	-596.56	&	$\ast$	\\
6	&	-409.53	&	 	&	-38.71	&	 	&	-425.36	&	$\ast$	\\
7	&	-227.94	&	 	&	-6.39	&	 	&	-251.63	&	$\ast$	\\
8	&	-193.42	&	 	&	23.66	&	 	&	-204.54	&	$\ast$	\\
9	&	-129.49	&	 	&	51.58	&	 	&	-132.82	&	$\ast$	\\
10	&	-84.46	&	$\ast$	&	57.75	&	 	&	-78.53	&	 	\\
11	&	-63.34	&	$\ast$	&	62.20	&	 	&	-51.41	&	 	\\
\hline
\hline
\end{tabular}
\label{aic_m_gls} 
\end{table}

\section{Conclusions}
In this paper, we critically examined two measures, phase-locking intervals (PLI) and global lability of synchronisation (GLS), proposed by Kitzbichler and colleagues \cite{bullmore} to characterise the presence of critical synchronisation in a system. We did so by presenting those measures with two very different models of synchronisation. In the first (Kuramoto Model) the oscillators are coupled with increasing $K$ to the mean field and undergo a critical transition. In the second (Independent Pairs Model) the oscillators are only allowed to couple in a pair wise manner. This latter model cannot be formulated as a system at criticality because there is no global coupling to associate the pairs with one another, and so no possibility of a mean field. 

When calculating the phase locking intervals (PLI) following the methodology of Kitzbichler et al. \cite{bullmore}, we showed that power laws were the best fit for a similar number of wavelet scales when considering PLI distributions for the critical, Kuramoto, model and the non-critical, Independent Pairs, model. The power law distribution and the slope found for the PLIs of the non-critical system was closely similar to that shown by the critical model. When further exploring the PLI probability distribution for coupling parameter values exceeding criticality, we found that the linearity of the log-log plot of the distribution at a number of wavelet scales still led to a best fit by a power law, suggesting that the observation of power laws within this framework can be present in a wide range of coupling values. We therefore conclude that the PLI measure should not be used to infer criticality (broadband or otherwise) in a system. 

In our simulations the GLS measure appeared better at discriminating between the critical, Kuramoto, system and the non-critical, Independent Pairs, model. We therefore conclude that GLS is a better measure than PLI for identifying critical systems, however, we believe that further work should be done to ascertain more precisely where its strengths lie, and compare it to other, non threshold-based methods such as proposed by Gong et al. \cite{gong}. In particular, we note that the GLS measure relies on counting the number of synchronised oscillators and that this depends crucially on how oscillators are defined, and distinguished. In the Kuramoto model, the number of oscillators is well defined, and each one is a discrete entity. With recorded neural activity, however, distinguishing multiple discrete oscillators is less straightforward. Kitzbichler et al., have applied the GLS measure to fMRI and MEG signals but its interpretation was limited by finite size effects (see loss of log-log linearity in the GLS distribution of MEG data in their figures 5D and 7D). To our knowledge the GLS measure has not been applied again to human neural data. Recently Meisel et al. \cite{meisel} have claimed to detect when compared to seizure-free electro-corticogram (ECoG) data a loss of adaptive self-organized criticality of the ECoG during epileptic seizures. This conclusion was arrived at through exploring power law scaling of ECoG phase locking using the PLI measure only. This is an exciting finding which received support from analysing the changes in PLI scaling seen in a computational model of self-organized criticality \cite{bornholdt}. However, our work indicates that interpreting the presence of a power law in the PLI probability distribution as a marker of criticality is problematic especially when a threshold has been applied to detect PLIs and when there has been pooling across many elements. 

\section*{Acknowledgements}
The authors would like to acknowledge: Dr M Kitzbichler for making his R code available, Dr J Cabral for providing her Matlab implementation of the Kuramoto model, Dr C Ginestet for useful discussions. MB was funded by CoMPLEX (Centre for Mathematics and Physics in the Life Sciences and Experimental Biology), University College London. SF was funded by UCLH CBRC (University College London Hospital, Comprehensive Biomedical Research Centre).  
All source code used in this study is available on request from the corresponding author.

\appendix

\section{Analytic Derivation of $\bigtriangleup_{ij}$}

The analytical solutions for $\bigtriangleup_{ij}$, the difference between phases $\theta_i$ and $\theta_j$ of oscillators $i$ and $j$, are distinct for the two cases $\frac{K}{\omega_i - \omega_j}>1$ and $\frac{K}{\omega_i - \omega_j}<1$ where $\omega_i$ and $\omega_j$ are the respective natural frequencies of oscillators $i$ and $j$, and $K$ is the coupling added globally to the system. We can rearrange Equation \ref{diffequation} to obtain the following integral:

\begin{equation*}\label{diff}
\int d t  = \int \frac{d\bigtriangleup}{(\omega_i - \omega_j) -K   \mbox{sin} (\bigtriangleup_{ij}) }
\end{equation*}
where $t$ denotes time.
This integral can be solved using the standard substitution of $x =  \mbox{tan} \left( \frac{\bigtriangleup_{ij}}{2} \right)$. 

Doing so, and letting $C = \frac{K}{(\omega_i - \omega_j)}$, we get:
\begin{align}\label{integral}
\int d t  & = \frac{2}{ (\omega_i - \omega_j)} \int{ \frac{ dx}{ \left(   1 - C^2 + \left(x - C \right)^2  \right)  }  }
\end{align}

There are two different scenarios for this integral, depending on whether $C<1$ and $\sqrt{1 - C^2}$ is a real or imaginary number. We deal with each case in turn. 

\subsection{If $C<1$, or when coupling is smaller than the difference in natural frequency}

\noindent We can rearrange \ref{integral} in terms of $\sqrt{1 - C^2}$ which is real and:
\begin{equation*}
\int d t  = \frac{2}{ (\omega_i - \omega_j)(1-C^2)} \int{ \frac{ dx}{ \left(   1 + \left(\frac{x - C}{\sqrt{1 - C^2}} \right)^2  \right)  } } 
\end{equation*}

We can solve this integral using the fact that $\mbox{tan}^{-1}(z) = \int \frac{dz}{1+z^2}$ to get:
\begin{align}\label{time}
\begin{split}
t  & = \frac{2}{ (\omega_i - \omega_j) \sqrt{(1-C^2)}} \left[  \mbox{tan}^{-1}\left( \frac{ \mbox{tan} \left( \frac{\bigtriangleup_{ij}}{2} \right) - C}{\sqrt{1 - C^2}}\right) \right. \\
       & \left. - \mbox{tan}^{-1}\left( \frac{ \mbox{tan} \left( \frac{\bigtriangleup^0_{ij}}{2} \right) - C}{\sqrt{1 - C^2}}\right)  \right] 
\end{split}
\end{align}

Here, $\bigtriangleup^0_{ij}$ is the value of  $\bigtriangleup_{ij}$ at time $t=0$, i.e., the initial difference in phase between oscillators $i$ and $j$. 

Setting $ D = \frac{2}{ (\omega_i - \omega_j) \sqrt{(1-C^2)}}  \mbox{tan}^{-1}\left( \frac{ \mbox{tan} \left( \frac{\bigtriangleup^0_{ij}}{2} \right) - C}{\sqrt{1 - C^2}}\right)$ we can rearrange Equation \ref{time} to get:
\begin{align*}
\begin{split}
\bigtriangleup_{ij} &= 2 \mbox{tan}^{-1} \left(    \left( \sqrt{1 - C^2}\right)  \right.  \\
                   &\left. \mbox{tan} \left(  \frac{\left( t-D\right) (\omega_i - \omega_j) \sqrt{(1-C^2) } }{2}   \right)   +C  \right)  
\end{split}
\end{align*}

\subsection{If $C>1$, or when coupling is larger than the difference in natural frequency}

Here, $\sqrt{1 - C^2}$ is imaginary, so we rearrange \ref{integral} in terms of $\sqrt{C^2 - 1}$:

\begin{equation*}
\int d t  = \frac{2}{ (\omega_i - \omega_j)(1-C^2)} \int{ \frac{ dx}{ \left(   1 - \left(\frac{x - C}{\sqrt{ C^2 - 1}} \right)^2  \right)  } }
\end{equation*}

We can solve this integral using the fact that $\frac{1}{2} \left( \mbox{log}^{-1}(-z-1) - \mbox{log}^{-1}(z-1) \right) = \int \frac{dz}{1-z^2}$:
\begin{equation*}
t  = \frac{-1}{ (\omega_i - \omega_j) \sqrt{(C^2-1)}} \mbox{log} \left[ A \left( \frac{1+y}{1-y} \right) \right]  
\end{equation*}
where $A = \frac{1-y^0}{1+y^0}$ and $y^0$ is the value of $y$ at time $t=0$.

This can be rearranged to yield:
\begin{equation*}
\bigtriangleup_{ij} = 2 \mbox{tan}^{-1} \left[  \sqrt{C^2 - 1} \left( \frac{e^{-t (\omega_i - \omega_j) \sqrt{(C^2-1)}} - A}{ A+e^{-t (\omega_i - \omega_j) \sqrt{(C^2-1)}}} \right) + C  \right]  
\end{equation*}

\newpage

\bibliographystyle{apsrev4-1}	
\bibliography{biblio}		

\newpage

 \end{document}